\documentclass{article}
\usepackage{arxiv}

\usepackage[utf8]{inputenc} 
\usepackage[T1]{fontenc}    
\usepackage{hyperref}
\hypersetup{
    colorlinks=true,
    linkcolor=blue,      
    citecolor=blue,      
    urlcolor=blue        
}
\usepackage{url}            
\usepackage{amsfonts}       
\usepackage{nicefrac}       
\usepackage{microtype}      
\usepackage{lipsum}         
\usepackage{newunicodechar}

\usepackage{graphicx}
\usepackage{booktabs}       
\usepackage{tabularx}
\usepackage{threeparttable}
\usepackage{multirow}
\usepackage{amsmath}
\usepackage{caption}        
\usepackage{float}          
\usepackage{placeins}       

\usepackage[numbers]{natbib}
\setcitestyle{numbers,square}
\usepackage{doi}

\title{Perceptions of the Metaverse at the Peak of the Hype Cycle: A Cross-Sectional Study Among Turkish University Students}

\author{ \href{https://orcid.org/0009-0007-5760-1914}{\includegraphics[scale=0.06]{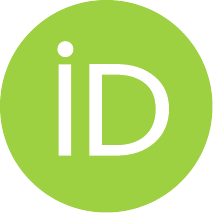}\hspace{1mm}Mehmet Ali~Erkan}\thanks{Data and analysis scripts are available at: \url{https://github.com/mrkn7/Metaverse-Perception-Study-2021}} \\
	Department of Statistics\\
	Middle East Technical University\\
	Ankara, Turkey \\
	\texttt{maerkan@metu.edu.tr} \\
	\And
	\href{https://orcid.org/0009-0002-7608-5218}{\includegraphics[scale=0.06]{orcid.pdf}\hspace{1mm}Halil Eren~Koçak} \\
    Department of Cognitive Science \\
	Middle East Technical University \\
	Ankara, Turkey \\
	\texttt{eren.kocak@metu.edu.tr} \\
}

\renewcommand{\shorttitle}{\textit{arXiv} Template}

\hypersetup{
pdftitle={A template for the arxiv style},
pdfsubject={q-bio.NC, q-bio.QM},
pdfauthor={ S.~Hippocampus, Elias D.~Striatum},
pdfkeywords={First keyword, Second keyword, More},
}

\begin{document}
\maketitle

\begin{abstract}
During the height of the hype in late 2021, the Metaverse drew more attention from around the world than ever before. It promised new ways to interact with people in three-dimensional digital spaces. This cross-sectional study investigates the attitudes, perceptions, and predictors of the willingness to engage with the Metaverse among 381 Turkish university students surveyed in December 2021. The study employs Fisher’s Exact Tests and binary logistic regression to assess the influence of demographic characteristics, prior digital experience, and perception-based factors. The results demonstrate that demographic factors, such as gender, educational attainment, faculty association, social media engagement, and previous virtual reality exposure, do not significantly forecast the propensity to participate in the Metaverse. Instead, the main things that affect people's intentions to adopt are how they see things. Belief in the Metaverse's capacity to revolutionize societal frameworks, especially human rights, surfaced as the most significant positive predictor of willingness. Conversely, apprehensions regarding psychological harm, framed as a possible “cyber syndrome,” represented a significant obstacle to participation.
Perceptions of technical compatibility and ethical considerations showed complex effects, showing that optimism, uncertainty, and indifference affect willingness in different ways. In general, the results show that early adoption of the Metaverse is based on how people see it, not on their demographics. The research establishes a historically informed benchmark of user skepticism and prudent assessment during the advent of Web 3.0, underscoring the necessity of addressing collective psychological, ethical, and normative issues to promote future engagement.
\end{abstract}


\section{Introduction}
Neal Stephenson first used the word "Metaverse" in his 1992 novel Snow Crash to describe a possible version of the Internet that would support permanent online 3-D virtual worlds \cite{stephenson1992snow}. The idea has changed over the past 30 years, but in late 2021, there was a lot of global interest and investment in it. This was after major rebranding efforts in the industry and improvements in virtual (VR) and augmented reality (AR) technology \cite{dwivedi2022metaverse}.

Supporters say that the Metaverse is a real society that goes beyond physical limits and could promote diversity, fairness, and humanism. But since the technology is still being developed, it raises a lot of ethical and privacy concerns, cybersecurity risks, and corporate dominance issues \cite{wang2023survey, yasuda2025metaverse, lee2021all}. It is important to know how university students, who are part of the "digital native" generation, feel about these virtual worlds because they could be the first to use them and help build them in the future. This study seeks to record the perceptions of university students in Turkey at the peak of Metaverse discussions (December 2021). We specifically look into how well people know the Metaverse idea and how they feel about it. The effect of gender identity on fear and risk assessment in virtual settings. The connection between how people use social media now and how willing they are to move to the Metaverse. This research captures student sentiment at this specific juncture, providing valuable empirical data on the initial reception of the Metaverse concept and emphasizing the skepticism that accompanied the technological hype.

\begin{figure}[ht]
    \centering
    \includegraphics[width=0.90\textwidth]{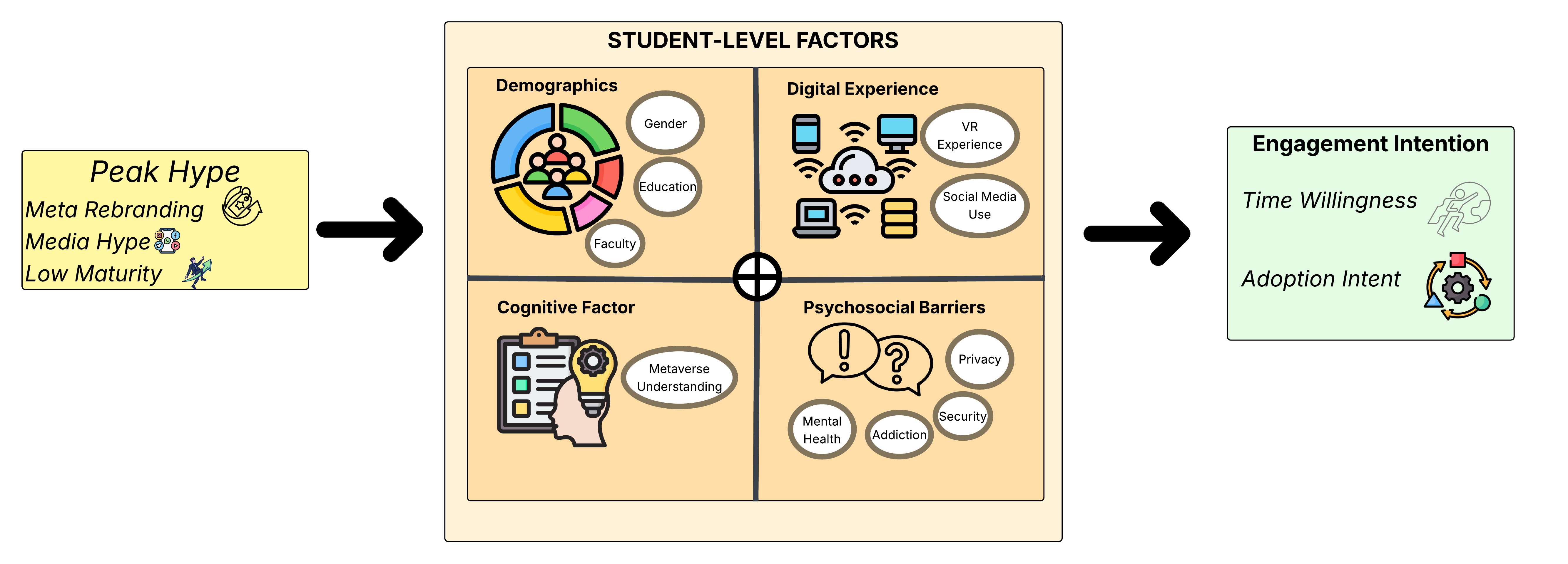}
    \caption{Conceptual Framework of Metaverse Engagement}
    \label{fig:framework}
\end{figure}

\section{Related Work}

The term "Metaverse" was first used in Neal Stephenson's 1992 science fiction novel, where it was a combination of "meta" (beyond) and "universe" \cite{stephenson1992snow}. In this groundbreaking book, Stephenson imagined a computer-made universe where people could interact as avatars in a 3D space that never ends. The terminology has been around for 30 years, but the technology to make it happen has only recently become possible \cite{encyclopedia2010031}. This is the next step in the evolution of the Internet, going from a 2D web of information to a 3D web of immersive experiences \cite{lee2021all}. The year 2021 marked a critical inflection point—often referred to as the "peak of the hype cycle"—precipitated largely by Facebook's rebranding to "Meta" in October 2021 \cite{dwivedi2022metaverse}. During this time, there was a lot of investment in industry and attention from around the world. This made the Metaverse seem like a "realistic society" where physical limits like geography could be overcome \cite{lee2021all}. But at the time, scholars warned that even though people were excited about the idea, it was still in its early stages of development and there was no unified scientific consensus or standard infrastructure \cite{encyclopedia2010031}. This timing is critical for understanding why student perceptions during late 2021 capture authentic early-stage adoption skepticism rather than mature technology assessment \citep{dwivedi2022metaverse}.

To understand how people adopt new technologies in this new world, we need strong theoretical foundations. The foundational Technology Acceptance Model (TAM) asserts that Perceived Usefulness (PU) and Perceived Ease of Use (PEOU) influence behavioral intention \citep{davis1989perceived}. TAM's relevance in immersive technologies has been rigorously evaluated, with recent research integrating both inhibiting and enabling factors, including cyber risks and perceived enjoyment \citep{sagnier2020user, al2023extending}. Research utilizing the Technology Acceptance Model (TAM) in clinical virtual reality contexts identified Perceived Usefulness as the singular significant predictor of the intention to use VR. This finding contests the notion that ease of use is universally essential, indicating that adoption drivers may vary by domain \citep{wallach2009virtual, sagnier2020user}. Moreover, the disparity between social media usage duration and Metaverse inclination noted in certain contexts indicates that conventional TAM frameworks may necessitate enhancement via theories such as the Theory of Planned Behavior, integrating psychological variables beyond mere utility assessments \citep{AJZEN1991179, jiang2025exploring}.

The main group of people who will use this is college students, who are the first real "digital natives"—people who grew up in digital environments \cite{prensky2006digital}. In Turkey, university students from Generation Z are described as being comfortable with using apps and websites and being native users of telecommunications technologies \citep{daim2024exploring}. Digital nativity does not ensure universal technology adoption; a study of Turkish university students' acceptance of information systems indicates that performance expectancy and facilitating conditions frequently surpass mere digital familiarity. Placing the study within Turkey's unique context facilitates interpretation \citep{ibrahimouglu2020digital}, as Turkish digital citizenship research indicates adoption patterns that differ from Western models, with infrastructure and support being significant factors \citep{ibrahimouglu2020digital, zickuhr2012digital}. To our knowledge, there has been no previous published research investigating the specific attitudes of Turkish university students toward the Metaverse during the peak-hype period, thus establishing the first empirical baseline for this significant demographic \citep{daim2024exploring}.

In this demographic, particular factors such as conceptual comprehension and previous experience are essential. Understanding how a technology works and what it can do may be necessary before someone decides to use it \citep{CAVAYE1995311}, which is in line with innovation diffusion theory, which says that perceived clarity predicts when people will start using something \citep{fishbein1977belief, CAVAYE1995311}. Consequently, conceptual ambiguity probably inhibited adoption intentions during the initial hype phase \citep{dwivedi2022metaverse, al2023extending}. Moreover, Virtual Reality (VR) experience constitutes a specific form of digital literacy, distinct from general technological proficiency \citep{rauschnabel2017adoption}. Prior experience with immersive technologies significantly impacts subsequent adoption, as immersive tendencies and absorption forecast both adoption and continued usage \citep{kober2013personality, herz2019understanding}.

On the other hand, significant barriers, such as differences based on gender and safety concerns, make adoption less likely. Historical research indicates enduring gender disparities, revealing that women frequently exhibit greater risk aversion towards technology adoption and demonstrate diminished optimism relative to men \cite{venkatesh2000don}. Gender does not consistently exert a direct influence on adoption intentions; however, it frequently moderates the predictive factors that inform decisions, with women being more swayed by subjective norms and perceived behavioral control \citep{henwood2008science, shaouf2018impact}. Recent studies indicate a convergence of adoption intentions across genders in virtual reality environments \citep{rauschnabel2017adoption}; however, gendered fear perceptions frequently endure \citep{wallach2009virtual}. These fears are made worse by a general lack of trust in digital security \citep{dwivedi2022metaverse}. Privacy concerns have been demonstrated to significantly impede adoption intentions, frequently superseding usability considerations \citep{morahan2000incidence, zickuhr2012digital}. Additionally, awareness of "cyber-syndrome"—physical, social, and mental disorders resulting from excessive internet use—illustrates the increasing global dialogue regarding technology's psychological externalities, serving as a significant impediment to the adoption of immersive environments \citep{turkle2015reclaiming, morahan2000incidence, dwivedi2022metaverse}.

\section{Methodology}
\label{sec:headings}

\subsection{Research Design and Sampling}

This study employed a cross-sectional survey design administered online via Google Forms across the period December 6-20, 2021.  The survey questionnaire was initially created in both Turkish and English to enhance accessibility; subsequently, all responses were standardized to Turkish due to the predominance of Turkish-speaking respondents. The target population consisted of undergraduate, master's, and doctoral students attending Turkish universities. Table\ref{tab:variables} shows the study variables, their descriptions, and the corresponding measurement scales.

The sampling method used was convenience snowball sampling, in which the first people to respond shared the survey link with their friends. This method, despite the risk of sampling bias, was feasible due to institutional limitations imposed by the pandemic and the nascent state of Metaverse discourse during the survey period. The first responses ($n = 398$) went through systematic data cleaning, which led to a final analytic sample of 381 observations after deleting records with a lot of missing data patterns.

\subsubsection{Sampling Bias Assessment Through Model Diagnostics}

Although convenience snowball sampling presents theoretical risks of homophily and network clustering, our empirical findings offer reassuring evidence of minimal sampling bias.

Initially, demographic variables (sex, education level, affiliation with the faculty) did not show significant correlations with willingness (all p > 0.05), accompanied by minimal effect sizes. If sampling bias had focused respondents within analogous demographic clusters, we would anticipate demographic characteristics to forecast outcomes. The lack of such patterns indicates that the sample effectively represented significant heterogeneity among university subpopulations.
Second, Table~\ref{tab:logistic_significant} shows the model's low pseudo-R² (0.226). It's mean that about 78\% of the variance in willingness is still not explained by the measured variables. This indicates that, although perception-based factors are essential, other individual differences—unrelated to sampling structure—also influence adoption intentions. If the sample had been significantly biased towards digitally-similar peers, we would anticipate that perception variables alone would account for a greater proportion of variance.
Third, model convergence happened without any problems with quasi-separation in most categories. This means that the response patterns were spread out and varied enough. 
This gives more proof that the sample isn't artificially concentrated.
These diagnostics do not completely remove concerns about sampling bias, but they give us reason to believe that the results are not just due to network homogeneity or severe selection bias.

\subsection{Statistical Analysis}

\subsubsection{Bivariate Analyses: Fisher's Exact Test and Cramér's V}

Associations between categorical variables and willingness to spend time in the Metaverse were examined using Fisher's Exact Test due to sparse contingency tables with expected cell frequencies less than 5. Fisher's Exact Test calculates the exact probability of the observed distribution using the hypergeometric distribution, making it particularly suitable for contingency tables with small expected frequencies, in contrast to Pearson's chi-square test which relies on large-sample approximations.

For a contingency table with cell frequencies $a, b, c, d$ and total sample size $n$, the probability $p$ of observing a specific configuration is given by:

\begin{equation}
    p = \frac{(a+b)!(c+d)!(a+c)!(b+d)!}{a!b!c!d!n!}
\end{equation}

Effect sizes for bivariate associations were quantified using Cramér's V, calculated as:

\begin{equation}
    V = \sqrt{\frac{\chi^2}{n \times \min(r-1, c-1)}}
\end{equation}

where $\chi^2$ is the chi-square statistic, $n$ is the total sample size, and $\min(r-1, c-1)$ is the minimum of rows minus one and columns minus one. Cramér's V ranges from 0 to 1, with values interpreted as follows: $V \leq 0.10$ (negligible), $0.10 < V \leq 0.20$ (weak), $0.20 < V \leq 0.40$ (moderate), and $V > 0.40$ (strong). Ninety-five percent confidence intervals for Cramér's V were calculated using the non-central chi-square distribution.

To verify the appropriateness of Fisher's Exact Test, we examined expected cell frequencies across all contingency tables. Perception variables (5 Likert-scale categories × 2 outcome levels) exhibited sparse data with 20-30\% of cells having expected frequencies less than 5. Demographic variables similarly showed sparse configurations. All expected frequency violations were within acceptable limits for Fisher's Exact Test application ($p < 0.05$ for sparse tables is valid with this test). Model convergence and coefficient stability were verified post-estimation.

\subsubsection{Multivariate Analysis: Binary Logistic Regression}

To identify independent predictors of willingness while controlling for potential confounders, we employed binary logistic regression with willingness to spend time in the Metaverse (High vs. Low/Neutral) as the dependent variable. Independent variables included perception-based factors (Metaverse ethics, human rights change expectations, compatibility concerns, cyber syndrome fears) and demographic characteristics (gender, education level).

The logistic regression model was specified as:

\begin{equation}
    \ln\left(\frac{P(Y=1|X)}{1-P(Y=1|X)}\right) = \beta_0 + \sum_{i=1}^{k} \beta_i X_i
\end{equation}

Parameters were estimated using Maximum Likelihood Estimation (MLE). Odds ratios with 95\% confidence intervals were calculated as $\text{OR}_j = e^{\beta_j}$ and $95\% \text{ CI}_{\text{OR}_j} = \left[\exp(\beta_j \pm 1.96 \times \text{SE}(\beta_j))\right]$.

Model fit was assessed using the Likelihood Ratio Test (LRT):

\begin{equation}
    \text{LRT} = -2[\ell_0 - \ell_1]
\end{equation}

where $\ell_0$ is the log-likelihood of the null model and $\ell_1$ is the log-likelihood of the fitted model. Additional fit indices included Nagelkerke's pseudo-$R^2$ and the Akaike Information Criterion (AIC).

High Willingness was defined as responses of 'Agree' (Katılıyorum) or 'Strongly Agree' (Kesinlikle Katılıyorum) on the 5-point Likert scale (Question 16: 'I would like  to spend most of my time in Metaverse'). All other responses ('Strongly Disagree', 'Disagree', 'Neither') were classified as Low/Neutral Willingness (0). The binary logistic regression design required dichotomization. To make sure the results were strong, a sensitivity analysis was done to look at different thresholds (like "Strongly Agree" only). The observations were independent because the data were gathered from a one-time cross-sectional survey in which each participant gave only one answer. The final analytical sample consisted of $N = 381$ respondents, including 33 events (8.7\%), corresponding to approximately 1.74 events per predictor after one-hot encoding of categorical variables. Although this ratio falls below the commonly recommended threshold of 10--20 events per predictor, it is considered acceptable in this context given the categorical nature of the predictors and the absence of substantial multicollinearity. Spearman rank-order correlations among perception variables ranged from $-0.12$ to $0.44$, indicating low to moderate associations well below the 0.70 threshold for problematic multicollinearity, and all variance inflation factors were below 2.0. Initial model estimation indicated quasi-separation in sparse categories, particularly for gender and compatibility-related variables; this issue was addressed by restricting gender to major categories (Male and Female) and aggregating compatibility issue categories, resulting in a reduced sample of $N = 375$ while maintaining representativeness. Following preprocessing, the model converged successfully under maximum likelihood estimation with no warnings related to non-convergence or numerical instability. Overall, diagnostic checks indicated that the model assumptions were reasonably satisfied, and the results are reportable with appropriate caution regarding the borderline events-to-predictors ratio.

\section{Data Analysis and Findings}
\label{sec:results}

The analysis of the 381 valid responses provided significant insights into the familiarity, risk perception, and governance preferences regarding the Metaverse among Turkish university students.

\subsection{Participant Characteristics}
Table~\ref{tab:demographics} presents the demographic characteristics of the participants. The final sample composition reflected: undergraduate students (69.6\%, n = 265), master's level students (20.2\%, n = 77), and doctoral students (10.2\%, n = 39). Faculty distribution encompassed Architecture, Arts and Sciences, Economic and Administrative Sciences, Education, Engineering, Law, Foreign Languages, and other academic units.

\begin{table}[htbp]
\centering
\begin{threeparttable}
\caption{Participant Demographics and Characteristics (N = 381)}
\label{tab:demographics}
\begin{tabular}{lccc}
\toprule
\textbf{Characteristic} & \textbf{n} & \textbf{\%} & \textbf{95\% CI} \\
\midrule

\textit{Gender Identity} \\
Male & 189 & 49.61 & [44.61--54.61] \\
Female & 186 & 48.82 & [43.84--53.82] \\
Prefer not to say & 6 & 1.57 & [0.72--3.39] \\

\addlinespace
\textit{Education Level} \\
Bachelor's & 265 & 69.55 & [64.76--73.96] \\
Master's & 77 & 20.21 & [16.48--24.53] \\
Doctorate & 39 & 10.24 & [7.58--13.69] \\

\addlinespace
\addlinespace
\textit{Faculty / College (Aggregated)} \\
Economics \& Administrative Sciences & 146 & 38.32 & [33.58--43.30] \\
Engineering & 75 & 19.69 & [16.00--23.97] \\
Health Sciences & 45 & 11.81 & [8.94--15.45] \\
Arts \& Sciences & 33 & 8.66 & [6.23--11.91] \\
Education & 32 & 8.40 & [6.01--11.62] \\
Law & 13 & 3.41 & [1.99--5.76] \\
Architecture \& Design & 10 & 2.62 & [1.42--4.78] \\
Communication & 9 & 2.36 & [1.25--4.41] \\
Other & 18 & 4.72 & [3.00--7.33] \\

\addlinespace
\textit{Prior Virtual Reality Experience} \\
Yes & 135 & 35.43 & [30.80--40.36] \\
No & 246 & 64.57 & [59.64--69.20] \\

\addlinespace
\textit{Social Media Usage (Daily Hours)} \\
$\leq$1 hour & 53 & 13.91 & [10.79--17.75] \\
2--3 hours & 194 & 50.92 & [45.91--55.90] \\
4--5 hours & 93 & 24.41 & [20.37--28.96] \\
6--7 hours & 34 & 8.92 & [6.46--12.21] \\
$\geq$8 hours & 7 & 1.84 & [0.89--3.74] \\

\addlinespace
\textit{Primary Social Media Platforms}\tnote{a} \\
WhatsApp & 368 & 96.59 & [94.25--98.00] \\
Instagram & 334 & 87.66 & [83.98--90.60] \\
YouTube & 323 & 84.78 & [80.82--88.04] \\
Twitter & 227 & 59.58 & [54.58--64.39] \\
Facebook & 142 & 37.27 & [32.56--42.23] \\
Telegram & 137 & 35.96 & [31.30--40.89] \\
LinkedIn & 111 & 29.13 & [24.80--33.89] \\
Snapchat & 87 & 22.83 & [18.90--27.31] \\
TikTok & 35 & 9.19 & [6.68--12.51] \\

\bottomrule
\end{tabular}

\begin{tablenotes}
\footnotesize
\item CI = Confidence Interval calculated using the Wilson score method.
\item[a] Multiple selections allowed; percentages do not sum to 100\%.
\end{tablenotes}
\end{threeparttable}
\end{table}

\subsection{Familiarity and Expectations}
\begin{figure}[H]
    \centering
    \begin{minipage}[t]{0.48\textwidth}
        \centering
        \includegraphics[width=\linewidth]{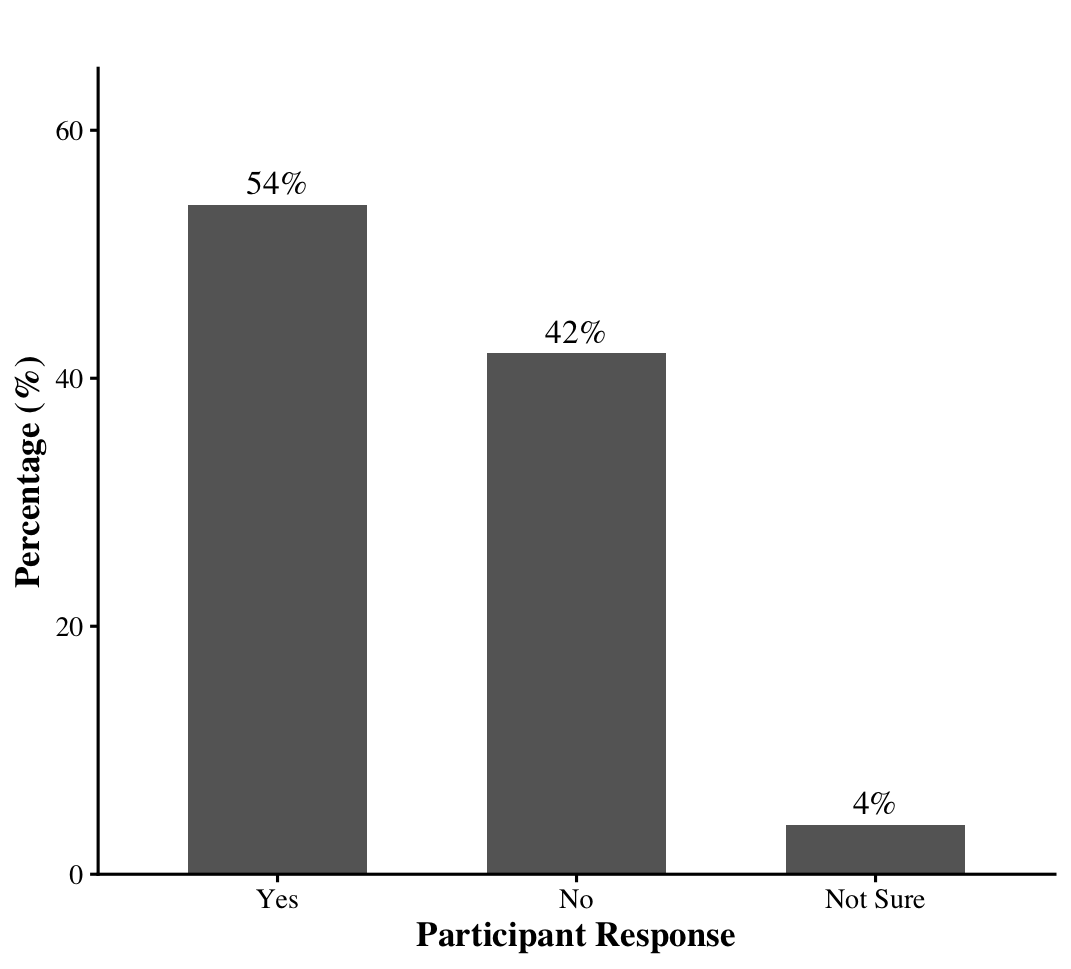}
        \caption{Participant Familiarity with Metaverse Concept}
        \label{fig:metaverse_concept}
    \end{minipage}
    \hfill 
    \begin{minipage}[t]{0.48\textwidth}
        \centering
        \includegraphics[width=\linewidth]{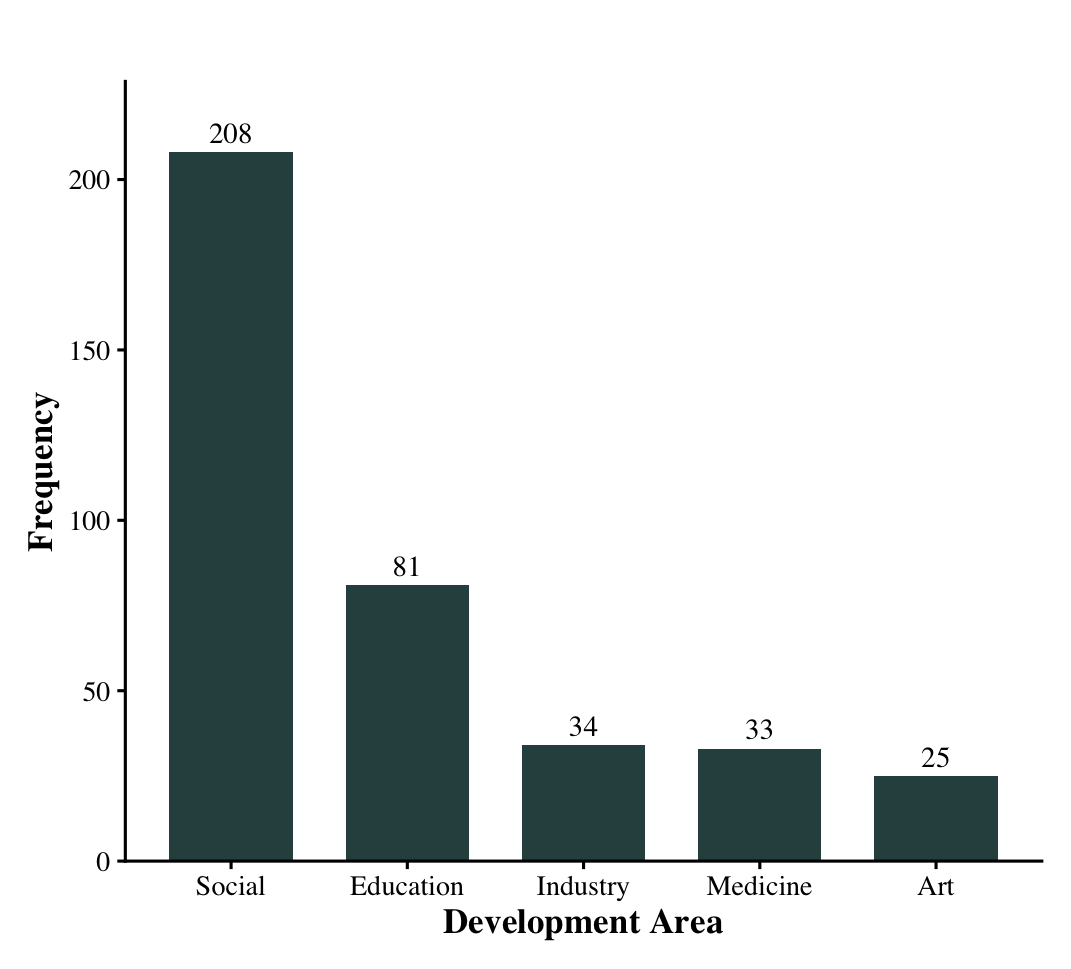}
        \caption{Perceived Domains of Impact of the Metaverse on Human Development}
        \label{fig:development_area}
    \end{minipage}
\end{figure}

Descriptive statistics revealed a moderate level of familiarity with the Metaverse concept during the peak hype period of late 2021. As shown in the Figure~\ref{fig:metaverse_concept}, 54\% of participants ($n=206$) stated they had heard of the concept before, while 42\% ($n=160$) were unfamiliar with it, and 4\% remained unsure. Then, regarding the anticipated impact of the Metaverse on human development the Figure~\ref{fig:development_area} shows that the majority of respondents identified the ``Social'' domain as the primary area for advancement ($n=208$), followed by Education ($n=81$), Industry ($n=34$), Medical ($n=33$), and Art ($n=25$). This suggests that despite the technological underpinnings, students primarily perceived the Metaverse as a sociological evolution rather than merely an industrial or artistic tool.

\subsection{Social Media Habits and Metaverse Adoption}
We investigated whether current digital engagement predicts the willingness to inhabit the Metaverse. According to Figure~\ref{fig:social_media_time}, participants' daily social media usage was categorized, with the largest group spending 2-3 hours daily ($n=194$).

\vspace{1em} 
\noindent 
\begin{minipage}[c]{0.40\textwidth} 
    \centering
    \includegraphics[width=\linewidth]{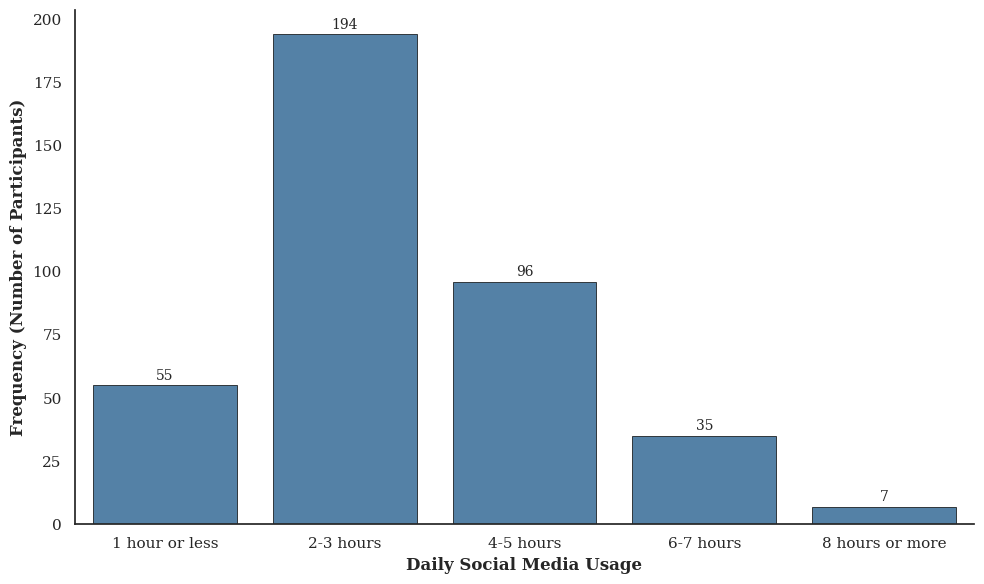} 
    \captionof{figure}{Daily Social Media Consumption.}
    \label{fig:social_media_time}
\end{minipage}%
\hfill 
\begin{minipage}[c]{0.58\textwidth}
    \centering
    \captionof{table}{Daily Social Media Usage vs. Willingness to Spend Time in Metaverse}
    \label{tab:social_vs_metaverse}
    
    \resizebox{\linewidth}{!}{%
        \setlength{\tabcolsep}{3pt} 
        \begin{tabular}{lcccccc}
        \toprule
         & \multicolumn{5}{c}{\textbf{Willingness to Spend Time in Metaverse}} & \\
        \cmidrule(lr){2-6}
        \textbf{Social Media Time} & \textbf{Str. Disagree} & \textbf{Disagree} & \textbf{Neither} & \textbf{Agree} & \textbf{Str. Agree} & \textbf{Total} \\ 
        \midrule
        \textbf{1 hour or less} & 24 & 16 & 9 & 3 & 1 & 53 \\
         & (45.3\%) & (30.2\%) & (17.0\%) & (5.7\%) & (1.9\%) & (100\%) \\ 
        \addlinespace
        \textbf{2-3 hours} & 55 & 63 & 60 & 11 & 5 & 194 \\
         & (28.4\%) & (32.5\%) & (30.9\%) & (5.7\%) & (2.6\%) & (100\%) \\ 
        \addlinespace
        \textbf{4-5 hours} & 25 & 36 & 24 & 3 & 5 & 93 \\
         & (26.9\%) & (38.7\%) & (25.8\%) & (3.2\%) & (5.4\%) & (100\%) \\ 
        \addlinespace
        \textbf{6-7 hours} & 7 & 10 & 12 & 4 & 1 & 34 \\
         & (20.6\%) & (29.4\%) & (35.3\%) & (11.8\%) & (2.9\%) & (100\%) \\ 
        \addlinespace
        \textbf{8 hours or more} & 1 & 4 & 2 & 0 & 0 & 7 \\
         & (14.3\%) & (57.1\%) & (28.6\%) & (0.0\%) & (0.0\%) & (100\%) \\ 
        \bottomrule
        \multicolumn{7}{r}{\textit{Note: Values represent Count (Row \%). $\chi^2=18.087$, $df=16$, $p=0.326$.}} \\
        \end{tabular}%
    }
\end{minipage}

\vspace{1em} 

Moreover, based on the table~\ref{tab:social_vs_metaverse}, high social media usage did not significantly predict willingness to spend time in the Metaverse ($p = 0.326$, Cramér's $V = 0.109$). Among respondents, both light users ($\leq$1 hour/day; 75.5\%, $n = 53$) and heavy users ($\geq$8 hours/day; 71.4\%, $n = 7$) demonstrated comparable levels of disagreement with spending the majority of time in the Metaverse. However, the small sample size of heavy users ($n = 7$) limits inference for this group. Fisher's Exact Test indicated no statistically significant association between daily social media consumption and Metaverse engagement intentions, suggesting that conventional digital engagement patterns do not predict early adoption.

\FloatBarrier

\subsection{Gender and Metaverse}

Fisher’s Exact Test was employed to examine whether students’ fear of potential negative outcomes in the Metaverse differs by gender identity. The results indicate a statistically significant association between gender identity and fear levels related to possible negativities in the Metaverse ($p = 0.008 < 0.05$). This finding suggests that gender identity plays a meaningful role in shaping students’ perceptions of risk in immersive digital environments.

\vspace{1em} 
\noindent
\begin{minipage}{\linewidth} 
    \centering
    \includegraphics[width=0.70\textwidth]{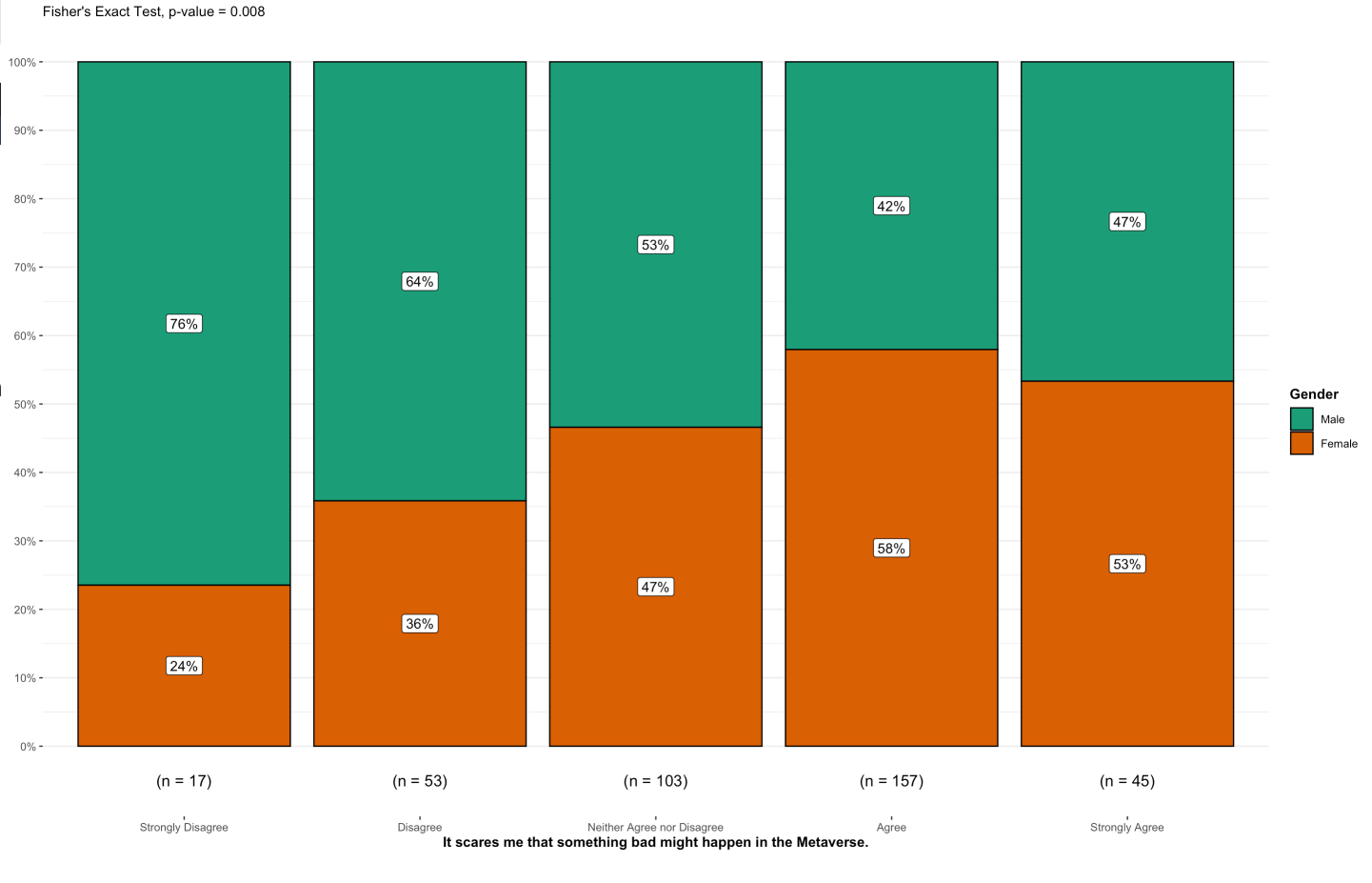}
    \captionof{figure}{Gender Differences in Fear Perception Regarding Metaverse Negative Outcomes.}
    \label{fig:gender_fear}
\end{minipage}

As illustrated in the figure~\ref{fig:gender_fear}, a substantial majority of respondents who selected ``Strongly Disagree'' were male (76\%), indicating that male students tend to report lower levels of fear regarding potential negative events in the Metaverse. In contrast, female respondents were more prevalent among those who answered ``Agree'' and ``Strongly Agree,'' reflecting higher levels of concern about possible risks. This distributional pattern supports the statistical test result and highlights a gender-based divergence in risk perception, with female students exhibiting greater sensitivity to potential negative consequences associated with Metaverse engagement.

\subsection{Governance Preferences and Institutional Trust}

Participants were asked who should establish the laws and rules in the Metaverse (Government, Companies, Both, or None). We analyzed how this preference correlates with ,Question 12, the belief that ``Laws, human rights, and the economy would change positively.'' Fisher's Exact Test revealed a significant relationship ($p = 0.0015$, Cramer's V = 0.149), indicating that one's preferred governance model is linked to their optimism regarding institutional evolution.

\begin{table}[ht]
\centering
\caption{Cross-tabulation: Governance Preference vs. Expectation of Positive Institutional Change}
\label{tab:governance_change}
\small 
\setlength{\tabcolsep}{3pt} 

\begin{tabular}{lcccccc}
\toprule
 & \multicolumn{5}{c}{\textbf{Expectation of Positive Change (Human Rights/Economy)}} & \\
\cmidrule(lr){2-6}
\textbf{Governance Pref.} & \textbf{Str. Disagree} & \textbf{Disagree} & \textbf{Neither} & \textbf{Agree} & \textbf{Str. Agree} & \textbf{Total} \\ 
\midrule
\textbf{Government} & 18 & 17 & 40 & 14 & 1 & 90 \\
 & (20.0\%) & (18.9\%) & (44.4\%) & (15.6\%) & (1.1\%) & (100\%) \\ 
\addlinespace
\textbf{Companies} & 3 & 14 & 37 & 13 & 6 & 73 \\
 & (4.1\%) & (19.2\%) & (50.7\%) & (17.8\%) & (8.2\%) & (100\%) \\ 
\addlinespace
\textbf{Both (Hybrid)} & 10 & 30 & 79 & 41 & 7 & 167 \\
 & (6.0\%) & (18.0\%) & (47.3\%) & (24.6\%) & (4.2\%) & (100\%) \\ 
\addlinespace
\textbf{None} & 7 & 10 & 19 & 6 & 2 & 44 \\
 & (15.9\%) & (22.7\%) & (43.2\%) & (13.6\%) & (4.5\%) & (100\%) \\ 
\bottomrule
\multicolumn{7}{r}{\parbox{12cm}{\raggedleft \textit{Note: Values represent Count (Row \%). Significant association found: Fisher's Exact Test, $p=0.0015$, Cramer's V = 0.149.}}} \\
\end{tabular}
\end{table}

The breakdown of responses offers nuanced insights into student expectations. As shown in Table~\ref{tab:governance_change}, the majority of respondents ($n=167$) favored a hybrid governance model involving both governments and companies. Notably, this group displayed the highest level of optimism, with approximately 29\% agreeing or strongly agreeing that human rights and laws would improve. This suggests that students view multi-stakeholder collaboration as the most viable mechanism for ensuring positive societal outcomes in the Metaverse.

In contrast, participants who advocated for exclusive government control ($n=90$) exhibited a paradoxical skepticism. While they preferred state regulation, nearly 39\% of this group disagreed or strongly disagreed that positive changes would occur. This finding implies a ``defensive'' preference; these students may view government intervention not as a guarantee of progress, but as a necessary buffer against potential corporate excesses, despite lacking confidence in the outcomes. Similarly, the ``None'' group, likely proponents of decentralization, mirrored this high skepticism (38.6\% disagreement), reflecting a distrust of all centralized authorities.

Across all categories, the modal response was ``Neither Agree nor Disagree'' (ranging from 43\% to 51\%). This pervasive uncertainty highlights that during the peak hype of late 2021, regardless of their governance preference, students struggled to envision exactly how legal and ethical frameworks would translate to virtual environments.

\begin{figure}[H]
	\centering
	\includegraphics[width=0.85\textwidth]{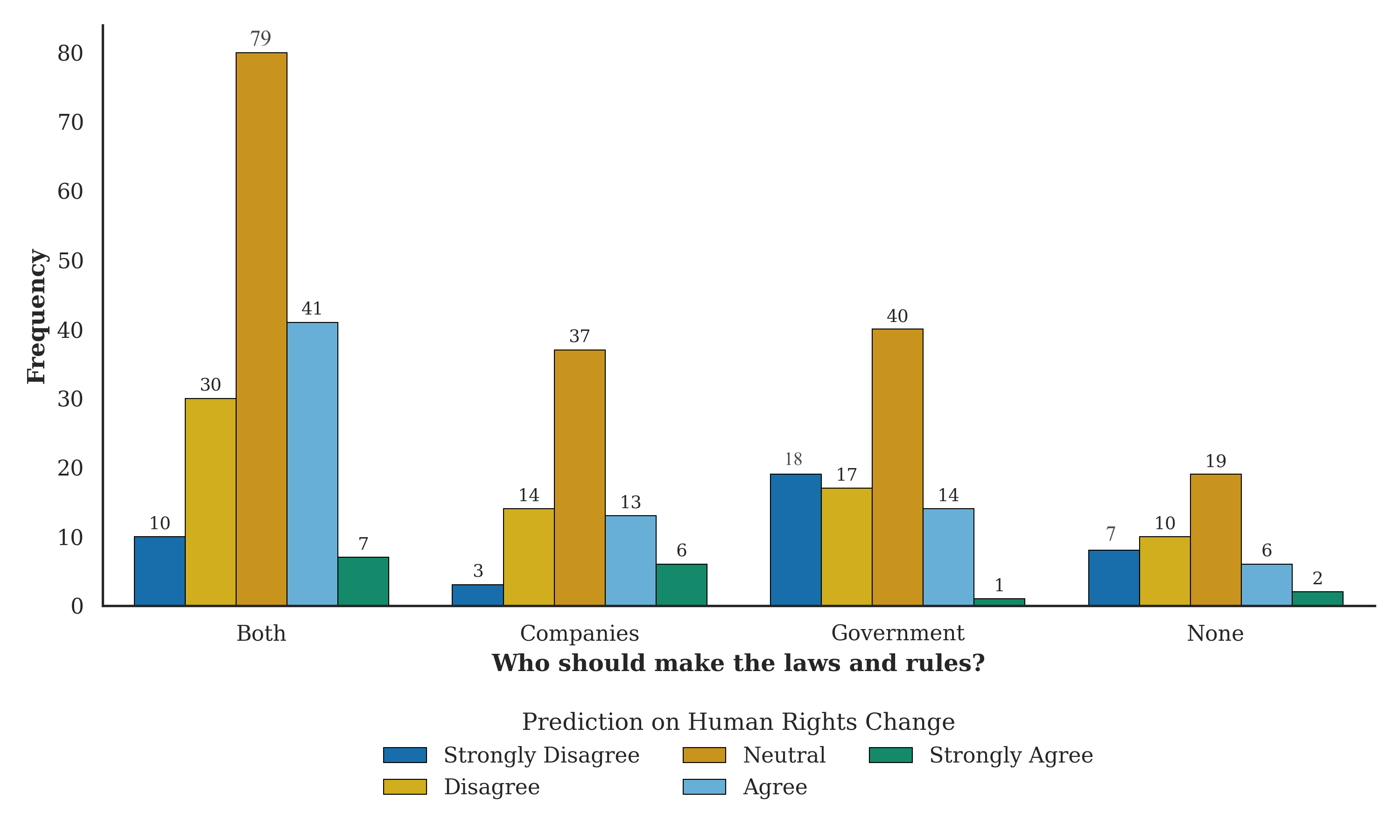}
	\caption{Associations between Governance Preferences and Expected Positive Institutional Change. The stacked bar chart presents response distributions across governance preference categories. Respondents preferring collaborative governance (Both) demonstrated the highest proportion of optimistic responses (green/blue segments). Conversely, the Government-only and None categories showed larger proportions of skepticism (dark blue/yellow segments), while uncertainty (Neither) remained the dominant sentiment across all groups.}
	\label{fig:governance_change}
\end{figure}



\subsection{Bivariate Associations with Willingness to Spend Time in Metaverse}

We conducted Fisher's Exact Tests to examine whether demographic characteristics and categorical variables predicted willingness to spend time in the Metaverse. Results are summarized in Table \ref{tab:bivariate_associations}.

\begin{table}[h!]
\centering
\caption{Bivariate Associations Between Demographics/Variables and Willingness to Spend Time in Metaverse (N=381)}
\label{tab:bivariate_associations}
\small
\begin{tabular}{|l|c|c|c|l|}
\hline
\textbf{Variable} & \textbf{p-value} & \textbf{Cramér's V} & \textbf{95\% CI} & \textbf{Significance} \\
\hline
Education Level & 0.7416 & 0.0396 & [0.00-0.13] & Not Significant \\
Faculty of Study & 0.1291 & -- & -- & Not Significant \\
Daily Social Media Time & 0.6699 & 0.0787 & [0.00-0.16] & Not Significant \\
Metaverse Concept Familiarity & 0.9165 & 0.0214 & [0.00-0.12] & Not Significant \\
Prior Virtual Experience & 0.7038 & 0.0157 & [0.00-0.10] & Not Significant \\
Gender & 0.8569 & 0.0025 & [0.00-0.09] & Not Significant \\
\hline
\multicolumn{5}{l}{\small Note: Effect size for Faculty not reported due to sparse contingency table across eight categories.} \\
\multicolumn{5}{l}{\small CI = Approximate 95\% confidence interval.} \\
\hline
\end{tabular}
\end{table}

For variables with sparse contingency tables (such as Faculty with eight categories), Cramér's V can yield inflated values that do not reflect true association strength. Following best practices in categorical data analysis, we report effect sizes only when contingency tables have adequate cell frequencies. For Faculty, the p-value is reported to indicate lack of statistical significance, while the effect size is omitted to avoid misleading interpretation.

Overall, none of the examined background and technology-related variables showed a statistically significant association with willingness to spend time in the Metaverse. Educational level, faculty of study, daily social media usage, prior virtual reality experience, gender, and self-reported familiarity with the Metaverse concept all yielded non-significant results with negligible effect sizes (Cramér’s V values close to zero). These findings suggest that Metaverse engagement intentions are largely independent of formal education, disciplinary background, conventional social media habits, prior exposure to immersive technologies, or basic conceptual awareness. Notably, even variables commonly assumed to predict adoption—such as VR experience and Metaverse familiarity—failed to differentiate willingness levels, indicating that surface-level exposure alone may be insufficient to drive engagement.

\subsection{Perception Variables and Willingness: Bivariate Analysis}

We examined associations between key perception variables and willingness using contingency table analysis and Fisher's Exact Tests.

\begin{table}[h!]
\centering
\caption{Perception Variables: Association with Willingness to Spend Time in Metaverse}
\label{tab:perception_bivariate}
\small
\begin{tabular}{|l|l|c|c|l|}
\hline
\textbf{Perception Variable} & \textbf{Category} & \textbf{High Willing} & \textbf{\% Willing} & \textbf{Notes} \\
\hline
\multirow{5}{*}{\textbf{Cyber Syndrome Concern}} & Strongly Disagree & 6/33 & 18.37\% & Reference \\
 & Disagree & 6/33 & 18.37\% & Similar to SD \\
 & Neutral & 2/37 & 5.43\% & Lower \\
 & Agree & 9/144 & 6.25\% & Lower \\
 & Strongly Agree & 9/230 & 3.91\% & Lowest \\
 & \textbf{Fisher's Exact Test} & \multicolumn{3}{c|}{$p = 0.0441$ (Significant)} \\
 & \textbf{Cramér's V} & \multicolumn{3}{c|}{0.149 [95\% CI: 0.06-0.22]} \\
\hline
\multirow{5}{*}{\textbf{Compatibility Issues}} & Strongly Disagree & 10/34 & 29.41\% & Highest \\
 & Disagree & 0/17 & 0.00\% & Polarized \\
 & Neutral & 6/45 & 13.33\% & Moderate \\
 & Agree & 13/212 & 6.13\% & Lower \\
 & Strongly Agree & 4/73 & 5.48\% & Low \\
 & \textbf{Fisher's Exact Test} & \multicolumn{3}{c|}{$p = 0.0010$ (Highly Significant)} \\
 & \textbf{Cramér's V} & \multicolumn{3}{c|}{0.246 [95\% CI: 0.14-0.31]} \\
\hline
\multirow{5}{*}{\textbf{Human Rights Change}} & Strongly Disagree & 0/51 & 0.00\% & Baseline \\
 & Disagree & 5/56 & 8.93\% & Low \\
 & Neutral & 6/128 & 4.69\% & Very Low \\
 & Agree & 7/72 & 9.72\% & Moderate \\
 & Strongly Agree & 14/34 & 41.18\% & Highest \\
 & \textbf{Fisher's Exact Test} & \multicolumn{3}{c|}{$p < 0.0001$ (Highly Significant)} \\
 & \textbf{Cramér's V} & \multicolumn{3}{c|}{0.358 [95\% CI: 0.27-0.43]} \\
\hline
\multirow{5}{*}{\textbf{Metaverse Ethics}} & Strongly Disagree & 3/17 & 17.65\% & Highest \\
 & Disagree & 6/36 & 16.67\% & High \\
 & Neutral & 2/81 & 2.47\% & Very Low \\
 & Agree & 12/166 & 7.23\% & Low \\
 & Strongly Agree & 10/81 & 12.35\% & Moderate \\
 & \textbf{Fisher's Exact Test} & \multicolumn{3}{c|}{$p = 0.0063$ (Significant)} \\
 & \textbf{Cramér's V} & \multicolumn{3}{c|}{0.174 [95\% CI: 0.08-0.25]} \\
\hline
\multirow{5}{*}{\textbf{Socioeconomic Impact}} & Strongly Disagree & 3/24 & 12.50\% & Moderate \\
 & Disagree & 5/49 & 10.20\% & Moderate \\
 & Neutral & 10/118 & 8.47\% & Low \\
 & Agree & 9/145 & 6.21\% & Low \\
 & Strongly Agree & 6/45 & 13.33\% & Moderate \\
 & \textbf{Fisher's Exact Test} & \multicolumn{3}{c|}{$p = 0.2470$ (Not Significant)} \\
 & \textbf{Cramér's V} & \multicolumn{3}{c|}{0.074 [95\% CI: 0.00-0.16]} \\
\hline
\end{tabular}
\end{table}


Unlike demographic and experiential variables, numerous attitudinal and expectation-based factors demonstrated robust and statistically significant correlations with the willingness to participate in the Metaverse. Beliefs about possible changes in human rights were the strongest predictor. People who thought these changes would happen were much more willing to accept them, showing a deep divide in worldviews between those who are hopeful about change and those who are skeptical. Perceived technical compatibility issues exhibited a significant and polarized correlation, indicating a threshold effect wherein heightened technological optimism correlates with maximal adoption intentions, while moderate concern aligns with minimal willingness. Psychological risk perceptions, especially fears associated with cyber syndrome, served as a substantial deterrent, demonstrating that apprehensions regarding mental well-being significantly inhibit engagement intentions. Ethical perceptions exhibited a moderate, non-linear relationship, indicating that individuals with strong opinions—regardless of their orientation—demonstrated greater willingness than those expressing neutrality. This suggests that ambivalence may be more obstructive than ethical skepticism. In contrast, perceptions of socioeconomic access barriers did not substantially affect adoption intentions, indicating that individual readiness to participate in the Metaverse is primarily influenced by normative, ethical, and psychological expectations rather than apprehensions regarding economic inequality.

\subsection{Multivariate Logistic Regression: Predictors of Willingness}

To identify independent predictors of willingness while controlling for confounding variables, we conducted binary logistic regression with the dependent variable (High Willingness vs. Low/Neutral) predicted by perception variables and demographics.

The logistic regression model was specified as:

\begin{equation}
\ln\left(\frac{P(\text{High Willingness})}{1-P(\text{High Willingness})}\right) = \beta_0 + \sum_{i=1}^{k} \beta_i X_i
\end{equation}

where $X_i$ included perception variables (ethics, human rights change, compatibility issues, cyber syndrome) and demographics (gender, education level).

\begin{table}[h!]
\centering
\caption{Binary Logistic Regression: Significant Predictors of Willingness to Spend Time in Metaverse}
\label{tab:logistic_significant}
\footnotesize
\begin{tabular}{|l|c|c|c|c|c|}
\hline
\textbf{Predictor} & \textbf{Coefficient} & \textbf{Std. Error} & \textbf{z-statistic} & \textbf{p-value} & \textbf{Odds Ratio (95\% CI)} \\
\hline
\multicolumn{6}{|l|}{\textit{\textbf{A. Significant Predictors (p < 0.05)}}} \\
\hline
Human Rights: Strongly Agree & 2.053 & 0.889 & 2.311 & 0.021* & 7.79 [1.37--44.48] \\
Cyber Syndrome: Strongly Agree & -1.884 & 0.877 & -2.148 & 0.032* & 0.15 [0.03--0.85] \\
Compatibility: Neutral & 1.396 & 0.658 & 2.123 & 0.034* & 4.04 [1.11--14.66] \\
Metaverse Ethics: Neutral & -2.182 & 0.779 & -2.802 & 0.005** & 0.11 [0.02--0.52] \\
\hline
\multicolumn{6}{|l|}{\textit{\textbf{B. Non-Significant but Substantively Notable Predictors}}} \\

 \\
\hline
Compatibility: Strongly Agree & 1.227 & 0.804 & 1.526 & 0.127 & 3.41 [0.71--16.50] \\
Cyber Syndrome: Strongly Disagree & -0.573 & 1.063 & -0.539 & 0.590 & 0.56 [0.07--4.53] \\
\hline
\multicolumn{6}{|l|}{\textit{\textbf{C. Non-Significant Predictors (p > 0.05)}}} \\
\hline
Gender: Female & 0.241 & 0.448 & 0.538 & 0.591 & 1.27 [0.53--3.06] \\
Education: Bachelor's & 0.232 & 0.685 & 0.338 & 0.735 & 1.26 [0.33--4.83] \\
Education: Master's & 0.569 & 0.751 & 0.757 & 0.449 & 1.77 [0.41--7.71] \\
Metaverse Ethics: Agree & -0.066 & 0.656 & -0.100 & 0.920 & 0.94 [0.26--3.39] \\
Metaverse Ethics: Strongly Disagree & -0.617 & 1.257 & -0.491 & 0.623 & 0.54 [0.05--6.34] \\
Metaverse Ethics: Strongly Agree & 0.679 & 0.845 & 0.804 & 0.421 & 1.97 [0.38--10.33] \\
Human Rights: Agree & 0.156 & 0.709 & 0.220 & 0.826 & 1.17 [0.29--4.69] \\
Human Rights: Neutral & 0.217 & 0.657 & 0.330 & 0.741 & 1.24 [0.34--4.50] \\
Compatibility: Strongly Disagree & 0.447 & 0.686 & 0.652 & 0.515 & 1.56 [0.41--6.00] \\
Human Rights: Strongly Disagree & -0.923 & 1.251 & -0.738 & 0.460 & 0.40 [0.03--4.61] \\
Cyber Syndrome: Neutral & -0.874 & 0.688 & -1.271 & 0.204 & 0.42 [0.11--1.61] \\
Cyber Syndrome: Agree & -0.107 & 0.536 & -0.200 & 0.841 & 0.90 [0.31--2.57] \\
\hline
\multicolumn{6}{|l|}{\small \textbf{Model Statistics:} $N=375$; Pseudo $R^2=0.226$; LLR $p<0.001$; LL$=-86.453$} \\
\hline
\end{tabular}
\end{table}

Logistic regression analyses revealed that normative and psychological beliefs play a decisive role in shaping Metaverse adoption intentions. Strong agreement with the view that human rights will fundamentally change in the Metaverse substantially increased willingness, making such respondents nearly eight times more likely to express high engagement compared to the baseline, underscoring the role of transformative optimism in adoption behavior. In contrast, strong concern about cyber syndrome and potential psychological harms markedly reduced adoption likelihood, with an 85\% decrease in odds, highlighting psychological safety as a major barrier to engagement. Attitudes toward technological compatibility further demonstrated a nuanced pattern: respondents expressing neutrality regarding future compatibility issues exhibited higher willingness than those holding strong positive or negative views, suggesting that openness to exploration may outweigh certainty-driven attitudes. Lastly, ethical perceptions displayed a counterintuitive but robust effect, whereby neutrality regarding Metaverse ethics was associated with the lowest likelihood of adoption—lower even than explicit ethical concern—indicating that disengagement from ethical discourse may reflect broader motivational indifference toward the technology, whereas strong ethical stances, regardless of direction, align with higher engagement intentions.

\section{Discussion}
\label{sec:discussion}

This study offers a systematic examination of university students’ willingness
to engage with the Metaverse during the peak of public attention in late 2021.
By jointly considering demographic factors, perception variables, and
multivariate modeling, the findings provide several theoretically and
practically relevant insights into early-stage Metaverse adoption.

A central finding of this study is that willingness to spend time in the
Metaverse is not structured along demographic lines. Neither gender, education
level, faculty affiliation, prior virtual reality experience, nor daily social
media usage significantly predicted adoption intentions. This demographic
independence was observed consistently across both bivariate analyses and the
multivariate logistic regression model.

These results challenge common assumptions in technology adoption research that
position younger, more educated, or more digitally experienced users as
automatic early adopters. Instead, the findings suggest that Metaverse adoption
at this early stage represents a population-wide hesitation rather than a
segmented diffusion process. The uniformly low willingness rate (8.7\%)
indicates the presence of shared structural or perceptual barriers that operate
across demographic groups.

Among all examined factors, belief in the transformative impact of the
Metaverse on human rights emerged as the strongest and most consistent predictor
of willingness. Respondents who strongly endorsed the idea that human rights
would change or adapt within the Metaverse were substantially more willing to
engage, exhibiting both the highest willingness rate (41.18\%) in bivariate
analysis and the largest effect size in the multivariate model (OR = 7.79).

This finding suggests that early Metaverse engagement is closely tied to
normative and ideological orientations rather than instrumental considerations.
Participants who perceive the Metaverse as a space of societal restructuring—
rather than merely a technological extension of existing platforms—appear more
motivated to participate. This aligns with diffusion theories emphasizing the
role of value alignment and perceived societal relevance in the adoption of
radically new technologies.

In contrast to the strong positive effect of normative optimism, concerns about
psychological harm—captured through the concept of a potential “cyber
syndrome”—constitute a major deterrent to Metaverse engagement. Strong agreement
with cyber syndrome concerns was associated with a pronounced reduction in
willingness, reflected in both a near five-fold difference in bivariate
willingness rates and a substantial decrease in adoption odds in the logistic
regression model (OR = 0.15).

Perceptions regarding technical compatibility issues revealed a nuanced and
polarized relationship with willingness. Participants who strongly rejected the
existence of compatibility problems exhibited the highest willingness rates,
indicating that technological optimism functions as an adoption catalyst.
Conversely, respondents expressing moderate disagreement showed virtually no
willingness, while those who remained neutral demonstrated elevated odds of
engagement in the multivariate model.

This non-linear pattern suggests that uncertainty does not necessarily suppress
adoption. Instead, neutrality may reflect openness to experimentation, whereas
moderate skepticism appears more discouraging than either strong optimism or
strong concern. Such threshold effects highlight the importance of perceived
technical reliability and clarity in shaping early adoption decisions.

Ethical perceptions displayed a distinctive pattern in which neutrality—not
ethical concern—was associated with the lowest willingness to engage. While
respondents holding strong ethical positions (either positive or negative)
exhibited moderate to high willingness, those expressing indifference toward
ethical issues were significantly less likely to adopt the Metaverse (OR =
0.11). This finding suggests that ethical disengagement may signal a broader lack of
motivation or interest in the technology itself. Rather than deterring
participation, ethical reflection—regardless of its direction—appears to be
characteristic of more engaged and attentive users. This underscores the role
of ethical discourse not as a barrier, but as a marker of involvement.

\section{Limitations and Future Directions}

This study offers significant insights into early perceptions of the Metaverse during a pivotal time of increased public scrutiny; however, several limitations must be acknowledged in the interpretation of the results.

\subsection{Methodological Constraints}

The study employs a cross-sectional snapshot design, documenting attitudes at a singular moment during the peak hype phase of late 2021. Consequently, causal relationships between perceptions and willingness to engage cannot be deduced. Longitudinal designs that monitor the evolution of perceptions, fears, and normative beliefs as the Metaverse develops would yield more robust evidence concerning the formation of attitudes and adoption trajectories.

\subsubsection{Sampling Bias and Representativeness}

Theoretically, convenience snowball sampling introduces selection bias, but various model-based diagnostics indicate that the sample reflects significant diversity. Specifically: (1) demographic variables exhibited no significant correlations with willingness, contesting the notion of severe network clustering; (2) the limited explanatory capacity of the logistic model suggests that unmeasured individual factors—not sampling structure—are responsible for the majority of variance in adoption intentions; and (3) the logistic regression model converged without quasi-separation complications, indicating sufficient response dispersion. These findings empirically confirm that the results are not significantly affected by sampling bias.

\subsection{Construct and Measurement Considerations}

At the time of data collection, the Metaverse was still a new and not very well-defined idea. Participants may have possessed diverse interpretations, spanning from gaming-centric virtual worlds to expansive concepts of immersive digital societies. Consequently, the familiarity measure ("Have you heard of the Metaverse?") assesses awareness but fails to differentiate between superficial recognition and profound conceptual comprehension. This may partially elucidate why conceptual familiarity did not forecast willingness, whereas more significant perceptual variables did.

Moreover, although the study included several key perception variables, other potentially
relevant factors—such as age, intensity of gaming experience, personality traits,
or broader technology readiness—were not measured. These variables may help
explain additional heterogeneity in willingness that remains unexplained by the
current model, which accounted for approximately 22.6\% of variance.

\subsection{Cultural and Contextual Scope}

The findings are situated within the Turkish higher education context and may
not generalize to other cultural or institutional settings. Differences in
digital infrastructure, institutional trust, regulatory expectations, and
normative beliefs about technology may shape Metaverse perceptions differently
across countries. Cross-cultural replications are therefore necessary to assess
the universality of the perception-driven adoption patterns identified in this
study.

\subsection{Future Research Directions}

Building on the present findings, future research should prioritize:
(1) probability-based sampling across diverse institutional and demographic
contexts; (2) longitudinal designs examining how willingness evolves as the
Metaverse becomes more tangible; (3) mixed-methods approaches integrating
qualitative insights into users’ normative beliefs and fears; (4) experimental
studies testing whether improvements in psychological safety, ethical
transparency, or technical reliability can causally increase willingness; and
(5) cross-cultural comparisons to identify context-dependent versus universal
drivers of early adoption.

Despite these limitations, the study is an important historical reference point for understanding how people first reacted to the Metaverse. The results demonstrate that adoption intentions are primarily shaped by perceptions and normative beliefs rather than demographic factors, thereby providing a foundation for subsequent theoretical and empirical investigations into immersive technology acceptance.

\section{Conclusion}
\label{sec:conclusion}

This study examined university students’ willingness to engage with the
Metaverse at a historically significant moment—the peak of public attention in
late 2021. Despite widespread awareness of the concept, the findings demonstrate
that heightened visibility and hype did not translate into strong adoption
intentions. Overall willingness to spend time in the Metaverse remained low,
indicating a cautious and evaluative stance rather than uncritical enthusiasm.

The results show that adoption intentions are not shaped by demographic
characteristics or prior digital experience. Gender, education level, faculty
affiliation, use of social media, and virtual reality exposure did not
significantly predict willingness to engage. Instead, willingness was driven
primarily by perception-based factors, highlighting the central role of
psychological, normative, and evaluative processes in early Metaverse reception.

Belief in the potential of the Metaverse to transform societal structures—particularly
human rights—emerged as the strongest positive predictor of willingness. In
contrast, concerns about psychological harm, conceptualized as a possible
"cyber syndrome,” constituted a substantial barrier to participation. Perceptions
of technical compatibility and ethical considerations further revealed nuanced
effects, suggesting that optimism, uncertainty, and indifference operate
differently in shaping early adoption decisions.

These findings indicate that early Metaverse adoption is best
understood as a perception-driven process rather than a demographically segmented
one. Students do not reject the Metaverse outright; rather, they assess it
through the lenses of safety, ethical meaning, and societal relevance. The
absence of demographic differentiation suggests that adoption barriers are
widely shared throughout the population, emphasizing the need for platform
developers and policymakers to address common concerns related to psychological
safety, ethical clarity, and tangible value.

By providing a systematic account of Metaverse perceptions during the peak hype
phase, this study establishes a valuable baseline for future research. As
immersive technologies evolve and move beyond speculative discourse, continued
empirical investigation will be essential to understanding whether—and under
what conditions—cautious evaluation gives way to sustained engagement.

\bibliographystyle{unsrtnat}
\bibliography{references}  

\appendix

\newpage

\section{Variable Definitions}
\label{app.variables}

\begin{table}[ht]
	\caption{Study Variables, Descriptions, and Measurement Scales}
	\label{tab:variables}
	\centering
	\small 
	\begin{tabularx}{\textwidth}{l >{\raggedright\arraybackslash}X >{\raggedright\arraybackslash}X}
		\toprule
		\multicolumn{3}{c}{\textbf{Survey Variables and Operationalization}} \\
		\cmidrule(r){1-3}
		\textbf{Variable} & \textbf{Description} & \textbf{Scale/Response Format} \\
		\midrule
		Gender & Gender identity of participant & Categorical (M/F/Other/Prefer not to say) \\
		Education Level & Current educational status & Categorical (Bachelor's/Master's/Doctorate) \\
		Faculty & Academic unit affiliation & Categorical (8 categories) \\
		Social Media Apps & Platforms used by participant & Multiple selection (Facebook, Instagram, etc.) \\
		Social Media Time & Daily hours spent on social media & Ordinal ($\leq$1 / 2-3 / 4-5 / 6-7 / $\geq$8 hours) \\
		\midrule
		Metaverse Concept & Familiarity with Metaverse term & Categorical (Yes/No/Unsure) \\
		Virtual Reality Exp. & Prior VR device experience & Categorical (Yes/No) \\
		Thoughts on Metaverse & General opinions about Metaverse & Categorical (5 options) \\
		Development Area & Anticipated domain of most development & Categorical (Education/Industry/Art/Medical/Social) \\
		Leading Company & Preferred corporate leader of Metaverse & Categorical (8 options) \\
		\midrule
		Laws \& Governance & Preferred creator of Metaverse rules & Categorical (Government/Companies/Both/None) \\
		Laws \& HR Change & Expected positive institutional change & Likert (1-5: Strongly Disagree to Strongly Agree) \\
		Achievement Satis. & Virtual achievement satisfaction vs. reality & Likert (1-5: Strongly Disagree to Strongly Agree) \\
		Fear of Metaverse & Fear of negative outcomes & Likert (1-5: Strongly Disagree to Strongly Agree) \\
		Ethical Concerns & Perception of Metaverse ethics & Likert (1-5: Strongly Disagree to Strongly Agree) \\
		\midrule
		Spending Time & Willingness to invest time in Metaverse & Likert (1-5: Strongly Disagree to Strongly Agree) \\
		Socioeconomic Factors & Impact of SES on Metaverse access & Likert (1-5: Strongly Disagree to Strongly Agree) \\
		Compatibility Issues & Expected technical compatibility problems & Likert (1-5: Strongly Disagree to Strongly Agree) \\
		Cyber-Syndrome & Expected long-term psychological effects & Likert (1-5: Strongly Disagree to Strongly Agree) \\
		\bottomrule
	\end{tabularx}
\end{table}

\FloatBarrier

\section{Survey Questionnaire}

The complete survey instrument administered to participants via Google Forms 
(December 6-20, 2021) is provided below.

\subsection{Demographics and Technology Use}

\textbf{Question 1.} What is your current gender identity?
\begin{itemize}
  \item[a)] Male
  \item[b)] Female  
  \item[c)] Other
  \item[d)] Prefer not to say
\end{itemize}

\textbf{Question 2.} What is your current level of education?
\begin{itemize}
  \item[a)] Bachelor's
  \item[b)] Master's
  \item[c)] Doctorate
\end{itemize}

\textbf{Question 3.} Which faculty does your department belong to?
\begin{itemize}
  \item[a)] Faculty of Architecture
  \item[b)] Faculty of Arts and Sciences
  \item[c)] Faculty of Economic and Administrative Sciences
  \item[d)] Faculty of Education
  \item[e)] Faculty of Engineering
  \item[f)] Faculty of Law
  \item[g)] School of Foreign Languages
  \item[h)] Other
\end{itemize}

\textbf{Question 4.} If you are using social media, please choose the 
applications that you use. (You can choose more than one)
\begin{itemize}
  \item[a)] Facebook
  \item[b)] YouTube  
  \item[c)] WhatsApp
  \item[d)] Instagram
  \item[e)] Twitter
  \item[f)] Telegram
  \item[g)] TikTok
  \item[h)] LinkedIn
  \item[i)] Snapchat
  \item[j)] Other
\end{itemize}

\textbf{Question 5.} How many hours do you spend on social media per day?
\begin{itemize}
  \item 1 or less
  \item 2-3
  \item 4-5
  \item 6-7
  \item 8 or more
\end{itemize}

\subsection{Metaverse Familiarity and Experience}

\textbf{Question 6.} Have you heard about Metaverse concept before?
\begin{itemize}
  \item Yes
  \item No
  \item I'm not sure
\end{itemize}

\textbf{Question 7.} Have you ever had virtual reality experience before 
(i.e., VR headset)?
\begin{itemize}
  \item Yes
  \item No
\end{itemize}

\textbf{Question 8.} Which of the following describe your thoughts about 
Metaverse?
\begin{itemize}
  \item I think Metaverse is a place that companies advertise their brands
  \item I would like to spend some time exploring the Metaverse
  \item I have no need for the Metaverse in my (real or virtual) life
  \item I still don't understand what the Metaverse is
  \item Other
\end{itemize}

\textbf{Question 9.} If the Metaverse spreads to the whole of our lives, in 
which areas do you believe human beings will develop more?
\begin{itemize}
  \item Education
  \item Industry
  \item Art
  \item Medical
  \item Social
\end{itemize}

\textbf{Question 10.} Which company do you want to lead the Metaverse universe?
\begin{itemize}
  \item Facebook
  \item Amazon
  \item Epic Games
  \item Disney
  \item Snapchat
  \item Microsoft
  \item Alibaba
  \item Sony
  \item Other
\end{itemize}

\subsection{Governance and Attitudes (Likert Scale)}

\textbf{Question 11.} In the framework of the Metaverse, should the government 
or companies make the laws and rules?
\begin{itemize}
  \item Governments
  \item Companies
  \item Both of them
  \item None of them
  \item Other
\end{itemize}

\textit{For Questions 12-19, indicate your level of agreement using the 
following scale:}
\begin{itemize}
  \item Strongly disagree
  \item Disagree
  \item Neither agree nor disagree
  \item Agree
  \item Strongly agree
\end{itemize}

\textbf{Question 12.} I think laws \& human rights \& economy etc. would change 
in the Metaverse positively.

\textbf{Question 13.} My achievements in the Metaverse world make me as happy 
and satisfied as in the real world.\\

\textbf{Question 14.} It scares me that something bad might happen in the 
Metaverse.

\textbf{Question 15.} I think the Metaverse concept is ethic.

\textbf{Question 16.} I would like to spend most of my time in Metaverse.

\textbf{Question 17.} I think that socio-economic factors will affect a person 
who wants to join to a Metaverse.

\textbf{Question 18.} If the world's leading companies create different 
Metaverses, I think that there will be compatibility issues (For example, the 
equipments I use for Facebook Metaverse would not work well in Amazon Metaverse).

\textbf{Question 19.} I think I would suffer from cyber-syndrome in long-term if 
I join a Metaverse. (Cyber-syndrome is the physical, social, and mental 
disorders that affect the human being due to the excessive interaction with the 
internet)\\

\subsection{Data Collection Details}

\begin{itemize}
  \item \textbf{Survey period:} December 6-20, 2021
  \item \textbf{Distribution method:} Snowball sampling via student networks
  \item \textbf{Target population:} University students in Turkey
  \item \textbf{Language:} Turkish (English translation available upon request)
  \item \textbf{Anonymity:} No personally identifiable information collected
  \item \textbf{Total responses:} 398 initial, 381 after data cleaning
\end{itemize}

\end{document}